\documentclass[aps,prl,superscriptaddress,twocolumn,floatfix,nofootinbib,preprintnumbers]{revtex4}
\usepackage{amsmath,multirow,graphicx,tabularx,epsfig}

\newcommand\one{\leavevmode\hbox{\small1\normalsize\kern-.33em1}}

\newcommand{\gev}{{\ensuremath\rm GeV}}
\newcommand{\tev}{{\ensuremath\rm TeV}}

\newcommand{\ifb}{{\ensuremath\rm fb^{-1}}}

\def\slashchar#1{\setbox0=\hbox{$#1$}           
   \dimen0=\wd0                                 
   \setbox1=\hbox{/} \dimen1=\wd1               
   \ifdim\dimen0>\dimen1                        
      \rlap{\hbox to \dimen0{\hfil/\hfil}}      
      #1                                        
   \else                                        
      \rlap{\hbox to \dimen1{\hfil$#1$\hfil}}   
      /                                         
   \fi}

\def\eg{{\sl e.g.} \,}
\def\ie{{\sl i.e.} \,}
\def\etal{{\sl et al} \,}

\newcommand{\be}{\begin{eqnarray*}}
\newcommand{\ee}{\end{eqnarray*}}

\newcommand{\bee}{\begin{eqnarray}}
\newcommand{\eee}{\end{eqnarray}}
\newcommand{\beeq}{\begin{equation}}
\newcommand{\eeeq}{\end{equation}}

\begin{document}



\title{Measuring Higgs Couplings from LHC Data}

\author{Markus Klute}
\affiliation{Massachusetts Institute of Technology, Cambridge, US}

\author{R\'emi Lafaye}
\affiliation{LAPP, Universit\'e Savoie, IN2P3/CNRS, Annecy, France}

\author{Tilman Plehn}
\affiliation{Institut f\"ur Theoretische Physik, Universit\"at Heidelberg, Germany}

\author{Michael Rauch}
\affiliation{Institut f\"ur Theoretische Physik, Karlsruhe Institute of Technology (KIT), Germany}

\author{Dirk Zerwas}
\affiliation{LAL, IN2P3/CNRS, Orsay, France}

\begin{abstract}
 Following recent ATLAS and CMS publications we interpret the results of their Higgs searches
 in terms of Standard Model operators. For a Higgs mass of 125 GeV we determine several
 Higgs couplings from 2011 data and extrapolate the results towards
 different scenarios of LHC running. Even though our analysis is
 limited by low statistics we already derive meaningful constraints on
 modified Higgs sectors.
\end{abstract}

\maketitle


If a scalar Higgs boson~\cite{higgs} exists, it should soon be discovered 
by ATLAS~\cite{atlas} and CMS~\cite{cms}. This would 
finally complete the Standard Model of elementary particles at the electroweak scale.
Going well beyond an observation of spontaneous symmetry breaking it
would establish the fundamental concept of field theories in
general and non-abelian gauge theories in particular.  

A minimal Higgs sector with only one physical state predicts all
properties of the Higgs boson, except for its mass: all couplings to
other particles are proportional to their
masses~\cite{abdel,spirix,lecture}.  Modifications of any kind
typically alter this structure and modify the relative coupling
strengths to different fermions and gauge bosons.\smallskip

One particularly appealing aspect of Higgs physics at the LHC is that
it is sensitive to new physics orthogonally to direct
searches~\cite{bsm_review}. First, the dimension-five operators (D5)
coupling the Higgs boson to gluons and to photons determine the main
Higgs production channel and one of the most promising decay
channels~\cite{abdel,spirix,lecture}. New particles coupling to the
Higgs boson will contribute to these operators~\cite{g4}. As an
example, chiral fermions do not even decouple, so their contributions
to Higgs production strongly constrain the model parameters.

Second, new physics can couple to the Standard Model through a
renormalizable dimension-four operator: a Higgs portal~\cite{portal}.
Typical LHC effects of such a portal include universally reduced Higgs
couplings and Higgs decays to invisible particles.  Similarly,
strongly interacting models in general alter the Higgs couplings
reflecting the structure of the underlying theory. Both of these
peculiarities of the Higgs sector illustrate the high priority of a
general Higgs coupling
analysis~\cite{sfitter_higgs,duehrssen,others,rome1,rome2,grojean}.\medskip

\underline{{\sc SFitter} Higgs analyses} --- the setup for our
analysis follows Refs.~\cite{sfitter_higgs} and~\cite{sfitter}. Any
Higgs coupling to SM particles is parameterized as
\begin{alignat}{9}
g_{xxH} &\equiv g_x  = 
\left( 1 + \Delta_x \right) \;
g_x^\text{SM} \; .
\label{eq:delta}
\end{alignat}
Independent variations of $\Delta_\gamma$ and $\Delta_g$ can be
included. Using ratios modified by $(1 + \Delta_{x/y})$ can eventually
be useful to cancel theoretical or systematic uncertainties.

The operator form of all couplings is given by the Standard Model, \ie
we assume the Higgs to be a parity-even scalar. An observation in the 
$\gamma\gamma$ channel suggests that we do not have to consider
spin-one interpretations~\cite{lookalikes}. Alternative
gauge-invariant forms for example of the $WWH$ coupling ($W_{\mu \nu}
W^{\mu\nu} H$) will eventually be testable in weak boson fusion
(WBF)~\cite{wbf_coup}.

\begin{figure*}[t]
\includegraphics[width=0.38\textwidth]{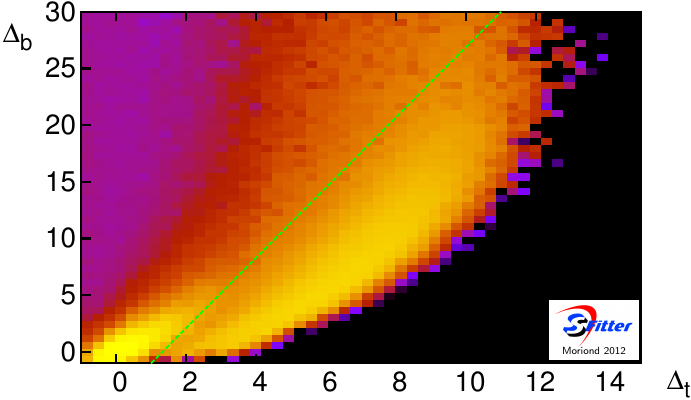}
\hspace*{0.05\textwidth}
\includegraphics[width=0.38\textwidth]{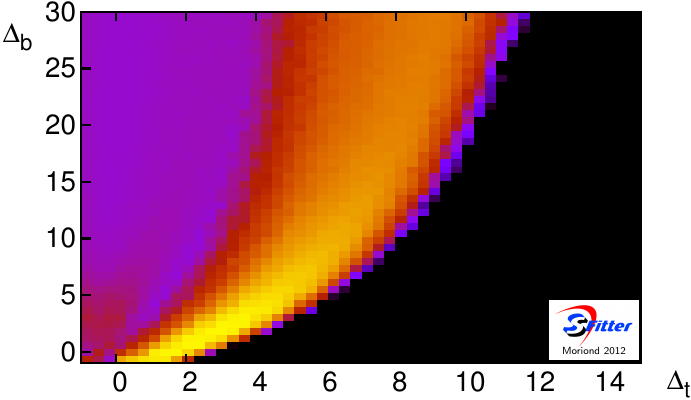}
\hspace*{0\textwidth}
\raisebox{-0.5pt}{\includegraphics[width=0.049\textwidth]{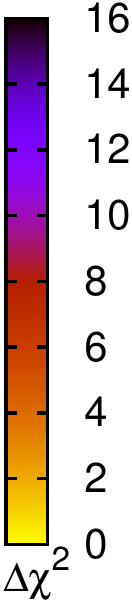}}
\vspace*{-3mm}
\caption{$\Delta_t$ vs $\Delta_b$ for the expected SM measurements
  (left) and the actual measurements (right), assuming $m_H =
  125$~GeV. The diagonal line separates the SM and the large-coupling
  solutions. For the actual data both solutions overlap.}
\label{fig:markov}
\end{figure*}

In the absence of a measurement of the Higgs width, which enters any
rate prediction, we assume~\cite{hdecay}
\begin{alignat}{9}
\Gamma_\text{tot} = \sum_\text{obs} \; \Gamma_x(g_x) 
+ \text{2nd generation} < 2\,\gev
\; ,
\label{eq:width}
\end{alignat}
the upper limit, corresponding to $\Delta_b \lesssim 28$, given by the
experimental resolution where width effects would be visible. 
The LHC will have no sensitivity to the $ccH$ coupling, which
contributes to the total width at the level of several
per-cent~\cite{abdel}; therefore, our additional model assumption
is linking the second-generation Yukawa couplings to their
third-generation counter parts, \eg $g_c = m_c/m_t \times
g_t^\text{SM} (1 + \Delta_t)$ with an appropriate scale choice in the
running masses.\smallskip

Beyond the available 2011 results~\cite{atlas,cms} we assume four scenarios: \\
\mbox{}\quad
\begin{tabular}{ll}
2011:                 & $(5~\ifb; 7~\tev)$ \\
2012${}_\text{low}$:  & $(7.5~\ifb, 8~\tev) \otimes (5~\ifb, 7~\tev)$ \\
2012${}_\text{high}$: & $(17.5~\ifb, 8~\tev) \otimes (5~\ifb, 7~\tev)$ \\
2014:                 & $(30~\ifb, 14~\tev)$ \\
HL-LHC:               & $(3000~\ifb, 14~\tev)$ 
\end{tabular}
\smallskip

We rely on fully correlated experimental and theoretical
uncertainties~\cite{gf_rate,wbf_rate,xs_group}.  This includes a
Poisson shape for counting rates and the centrally flat {\sc Rfit}
scheme for theory uncertainties~\cite{rfit}.  For the 7~TeV (and
8~TeV) run we use background rates, efficiencies and experimental
uncertainties as published by ATLAS and CMS~\cite{atlas,cms}.  For 14~TeV our input is 
described in Ref.~\cite{sfitter_higgs,duehrssen}.

Our analysis starts with a fully exclusive log-likelihood map of the
parameter space. Lower-dimensional distributions we project using a
profile likelihood. The best-fitting parameter points we identify
using {\sc Minuit}. Finally, the quoted 68\% confidence levels we
obtain from $5000$ toy measurements.\medskip

\underline{2011 fit to Standard Model} --- in a first attempt we use
2011 data to determine all couplings to heavy Standard Model
particles. The effective Higgs couplings to gluons and photons are
limited to SM loops.

To study general features of the Higgs parameter space with the help
of a global log-likelihood map we first assume a set of measurements with the SM expectation
as central values, but with the uncertainties of the
2011 data set. For ATLAS and CMS our 2011 data set includes all
$\gamma \gamma$, $ZZ$, $WW$, $\tau\tau$ and $b\bar{b}$ channels,
separated by the number of recoil jets, if available~\cite{atlas,
  cms}. In the $\gamma \gamma$ channel of CMS we separate the 2-jet
mode~\cite{cjv,manchester}. As in the published results
the remaining categories are combined.  A separation of the eight
inclusive channels into soft and hard $p_{T,H}$ might eventually be
beneficial; however, in 2011 weak boson fusion and $VH$ associated
production only contributed 10\% to 20\% to the rate, not giving 
measurable numbers of events~\cite{rome2}. \smallskip

\begin{figure}[b]
\includegraphics[width=0.42\textwidth]{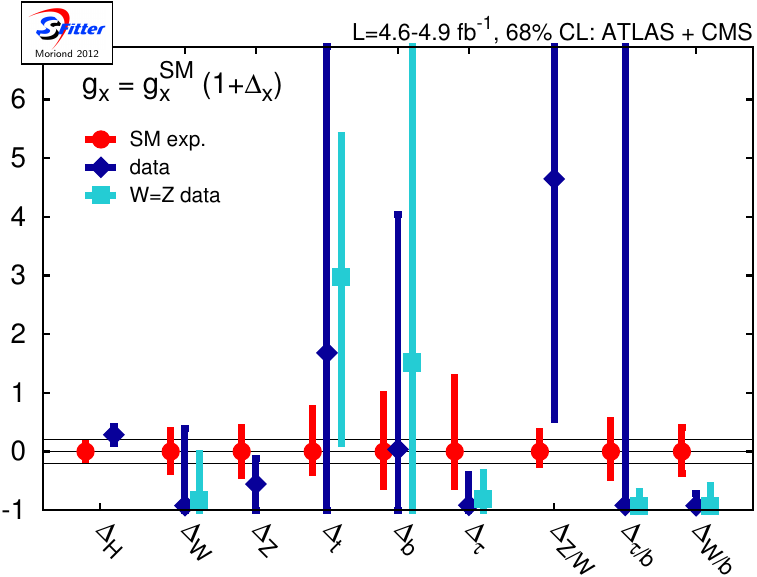}
\caption{Results with 2011 data, for the SM signal expectation and for the
  data ($m_H = 125$~GeV). For the latter we also show $\Delta_W
  = \Delta_Z$. The band indicates a $\pm 20\%$ variation.}
\label{fig:sm7}
\end{figure}

In the left panel of Fig.~\ref{fig:markov} we see that two scenarios
fit the expected Standard Model measurements: the SM-like solution
only allows for moderate values $\Delta_{b,t} \lesssim 3$. An
alternative large-coupling solution appears for a correlated increase
of $\Delta_t$ and $\Delta_b$ towards large values, with a best fit in
the $\Delta_{t,b} = 4-5$ range.  It reflects a cancellation between
the effective $ggH$ coupling ($\Delta_t$) and the total width. 
To reach this large-coupling regime from the SM-like
solution the $H \to \gamma\gamma$ rate has to stay stable, which
requires $\Delta_W$ to increase with $\Delta_t$. However, the $H \to
WW$ measurements do not allow for such a correlation. So when for
increasing $\Delta_b$ there is a point where $\Delta_W$ switches back to the SM
regime, $\Delta_t$ adjusts the large effective $\gamma \gamma H$
coupling, defining a secondary starting point at $\Delta_t \sim 3$
with a changed sign of $g_\gamma$.

For the expected SM central values we can separate the two solutions,
as indicated in Fig.~\ref{fig:markov}. In the absence of any
$t\bar{t}H$ rate measurement, \ie for the 7~TeV and 8~TeV runs, we
limit our extraction to the SM regime. This restriction is justified
by theory because top Yukawa couplings of $5 \times m_t = 875$~GeV
require a UV completion already at the scale of this Yukawa coupling.
Nevertheless, we have checked that enforcing the large-coupling regime
instead does not pose any technical problems. 

In the right panel of Fig.~\ref{fig:markov} we see that for the actual
measurements the two solutions are not separable. This is due to a
best-fit value around $g_W \sim 0$, so the effective $\gamma\gamma H$
coupling is always dominated by the top loop.\smallskip

In Fig.~\ref{fig:sm7} we show the error bars on the best fit values
from the 2011 run. Red dots correspond to the expected measurements,
fixing $\Delta_x = 0$ but including the correct uncertainties.  Typical error
bars for many couplings range around $\Delta_x = -0.5...1$,
corresponding to a variation of $g_x$ by a factor two. Forming ratios
slightly improves the results. Blue diamonds show the 2011
measurements. As mentioned before, the best fit resides around
$\Delta_W = -1$. Because we cannot ignore the large-coupling solution,
$\Delta_b$ and $\Delta_t$ now cover a significant correlated
enhancement. The best fit at $\Delta_\tau \sim -1$ reflects
inconclusive results.

Independently varying $\Delta_W$ and $\Delta_Z$ will typically lead to
a conflict with electroweak precision data. Because the measurements do
not include searches for new particles which might compensate for such
an offset~\cite{grojean}, we cannot include these constraints in our
fit. However, we can constrain our fit to $\Delta_W = \Delta_Z$. In
Fig.~\ref{fig:sm7} we see that this condition stabilizes the fit, and
we get a wide log-likelihood plateau at $\Delta_{W,Z}= -1...0$.
However, the large-coupling solution still overlaps with the SM
regime.\smallskip

After confirming that 2011 data has sensitivity to the individual
Higgs couplings we can simplify our hypothesis to arrive at tighter
constraints. The simplest hypothesis is a universal shift of
all Higgs couplings
\begin{equation}
  \Delta_x \equiv \Delta_H  \qquad \text{for all} \; x \; .
\end{equation}
This form factor could reflect mixing in a Higgs portal or the
strongly interacting nature of a composite Higgs. The first entry in
Fig.~\ref{fig:sm7} shows that the best fit of $\Delta_H$ to the 2011
data is at $\Delta_H = 0.28 \pm 0.14$, consistent with zero. 
Its expected and observed
error bars agree. This corresponds to the current
local significance of the Higgs hypothesis if we keep in mind that
$\sigma \times \text{BR}$ scales like $(1 + \Delta_H)^2$. The slightly
high value from data is an effect of the overlapping large-coupling
solution.\medskip

\begin{figure}[t]
\includegraphics[width=0.42\textwidth]{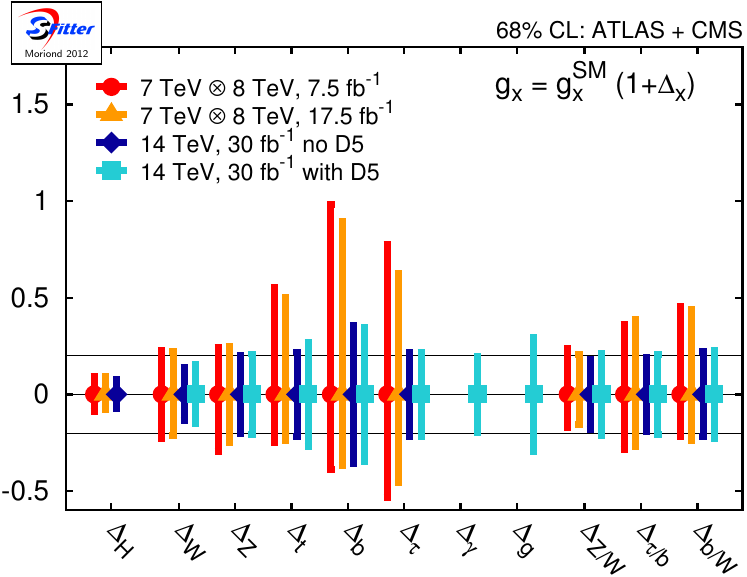}
\caption{Projections for an assumed SM signal at $m_H = 125$~GeV. The
  band indicates a $\pm 20\%$ variation.}
\label{fig:sm7814}
\end{figure}

\underline{Standard Model projections} --- in 2012 we expect Higgs
analyses to make major progress. Including a significant amount of
8~TeV data will increase the constraining power of WBF processes. Any
(close to) 14~TeV run should finally probe the top Yukawa directly, so
we can include $\Delta_\gamma$ and $\Delta_g$ as independent model
parameters.

The 2011 fit shows that the expected and the observed error bars on
the Higgs couplings are similar, but that the observed central values
lead to problems with a non-separable large-coupling solution. Because
the 2011 data will statistically not dominate the 2012 analysis, we
use expected measurements on the Standard Model values for all
projections.

For the 8~TeV results we use the same Higgs channels as have been
reported for the 2011 run at 7~TeV, with scaled-up rates and uncertainties.
The analysis for 14~TeV collider energy follows
Refs.~\cite{sfitter_higgs,duehrssen}. An additional observation of the
$t\bar{t}H, H \to b\bar{b}$ channel~\cite{bdrs,tth} could significantly bolster our
results.\smallskip

Fig.~\ref{fig:sm7814} shows the measurements we can expect from the
near future. Comparing the 2012 expectations with the 2011 results
shown in Fig.~\ref{fig:sm7} we see that $\Delta_{W,Z}$~\cite{wbf_w}
and $\Delta_\tau$~\cite{wbf_tau} benefit from enhanced WBF production
channels. The couplings to heavy quarks can only really be probed once
we include the full set of 14~TeV channels with $t\bar{t}H$ production
and $H \to b\bar{b}$ decay channels~\cite{bdrs,tth}. The direct
measurement of all SM Higgs couplings then allows us to not
only probe the structure of the Higgs sector but also search for new
physics effects in the effective couplings $g_g$ and $g_\gamma$. Error
bars in the 20\% range for both of these higher-dimensional operators
can strongly constrain any new particles which either rely on the
Higgs mechanism for their mass generation or couple to scalars like
the Higgs boson.

\begin{figure}[b]
\includegraphics[width=0.42\textwidth]{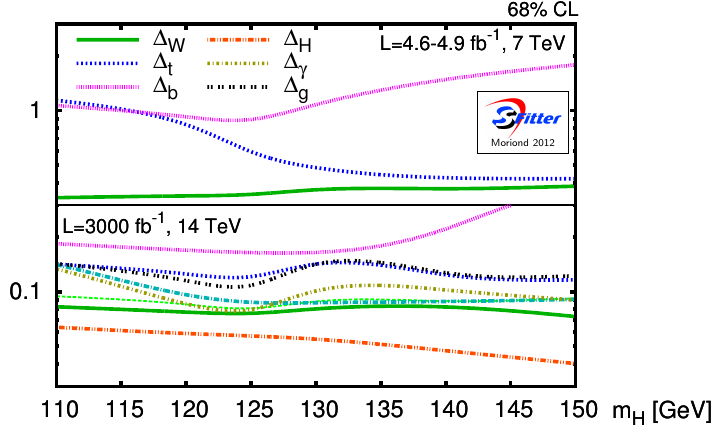}
\caption{Error bars for an assumed SM signal as a function of the
  Higgs mass for 2011 data (top) and the HL-LHC with 14~TeV and
  $3000~\ifb$ (bottom).}
\label{fig:mass}
\end{figure}

Finally, in Fig.~\ref{fig:mass} we show the dependence of some
expected error bars on the Higgs mass. Again, we assume Standard Model
measurements, so we only quote the error bar for the 2011 results and
for a very rough HL-LHC extrapolation. For the 2011 results we see that
$m_H = 125$~GeV is indeed a particularly lucky spot. Taken with the
appropriate grain of salt the HL-LHC projections show a very significant
improvement, but the naive statistics-dominated scaling with
luminosity does not apply any longer.\medskip

\underline{Exotic hypotheses} --- until now we have limited our fit to
the extraction of SM-like couplings. Given their good agreement with
2011 data and the lack of hints for new physics at the LHC, this
hypothesis is well motivated. Modest deviations from a Standard Model
Higgs sector include either supersymmetric or more general type-II
two-Higgs-doublet models, as well as form factors or mixing angles
affecting the Higgs couplings in a more or less constrained manner.

More exotic Higgs hypotheses illustrate the statistical limitations of
the 2011 measurements. Ignoring obvious problems with the UV
completion of such models we interpret the 2011 measurements in terms
of a fermiophobic and a gaugephobic Higgs model~\cite{gaugephobic},
\ie we assume that an observed 125~GeV resonance only couples to
gauge bosons or to fermions. For each model we compute the best
$\chi^2 = -2 \log L$ value:
\begin{center}
\begin{tabular}{l|r}
\hline
hypothesis      & $\chi^2_{2011}/\text{dof}$\\ \hline
independent $\Delta_x$     &  9.3/22 \\
$\Delta_W = \Delta_Z$      & 12.3/23 \\
$\Delta_W = \Delta_Z$ and $\Delta_b = \Delta_t = \Delta_\tau$ & 18.0/23 \\
$\Delta_x \equiv \Delta_H$ & 18.6/26 \\
gaugephobic                & 13.2/24 \\ 
fermiophobic               & 16.0/25 \\ \hline
\end{tabular}
\end{center}
The set of SM-like free couplings gives an excellent fit.
The good performance for a Higgs boson only coupling to fermions
is based on effective Higgs couplings to gluons and photons all
generated by heavy fermions. Only the $H \to ZZ$ channel implies a
slight statistical price to pay. 

The fermiophobic Higgs hypothesis is more of a challenge, because WBF
and $VH$ are the only production channels. All relevant branching
ratios grow as $g_b$ vanishes. In addition, $g_\gamma$ increases
without the destructive interference between the $W$ and top
loops. The best-fit point lies in the neighborhood of $\Delta_{W} \sim
-0.8$ and $\Delta_{Z} \sim 0$. For example in the $H \to ZZ$ channel
WBF and $ZH$ production are roughly equal and get combined with a
branching ratio around 75\%. The observed $WW$ channel still forces
$g_W$ to be small, so too few photon events are predicted for a
perfect fit. We note, however, that 2011 data
is too scarce to meaningfully distinguish between all these hypothesis.\medskip

\underline{Outlook} --- in this first comprehensive Higgs coupling
analysis we show that the published ATLAS and CMS measurements are
well explained by a Standard Model Higgs boson. The reason that the
2011 coupling measurements are weaker than the expected results is
a secondary large-coupling solution which cannot be separated. In
particular, a universal Higgs form factor $\Delta_H$ can be deduced
with a precision of 15\% and is in good agreement with unity .

In the future, major improvements of the Higgs coupling measurements 
will be an increased sensitivity to WBF production processes and
the direct measurement of the heavy quark Yukawas, \eg in $t\bar{t}H$
production. Unfortunately, the Higgs self coupling is still an
unsolved problem for $m_H \sim 125$~GeV~\cite{selfcoup}. 

The LHC projections presented in this paper might well serve as part
of the scientific case for a future Linear Collider Higgs
factory.\medskip

\underline{Acknowledgments} --- we are grateful to Peter Zerwas and
Dieter Zeppenfeld for their constant support of this analysis and to
Michael D\"uhrssen for helpful discussions. DZ and RL acknowledge the
useful discussions in the GDR Terascale (IN2P3/CNRS).  TP is grateful to
the 2011 Higgs Workshop in Eugene/OR, where many aspects of this
analysis were discussed in a very productive atmosphere.  Finally, we
would like to thank {\sc Fittino} for many years of pleasant physics
discussions.


\end{document}